\documentstyle[prl,aps,floats,psfig]{revtex}

\begin{document}

\widetext
\draft
\twocolumn[\hsize\textwidth\columnwidth\hsize\csname @twocolumnfalse\endcsname
\title{Comment on ``Diffusion of Ionic Particles in Charged Disordered Media''}

\author{
Michael W. Deem
}
\address{
Chemical Engineering Department, University of
California, Los Angeles, CA  90095-1592\\[0.1in]
}

\maketitle
]

In a recent Letter, Mehrabi and Sahimi discuss motion of ions
in charged disordered media, presenting
a variety of results obtained by Monte
Carlo simulation on a lattice \cite{Sahimi}.
Their observations of mean square displacements $R^2(t)$
suggest that this model exhibits anomalous diffusion
in three dimensions:
\begin{equation}
 R^2(t) \sim ({\mathrm const}) t^{1-\delta}~~ {\mathrm as}~t \to \infty \ .
\label{3a}
\end{equation}
They observe the same behavior in one and two dimensions, but do not
present results for $\delta$.  Mehrabi and Sahimi also make the
physically-surprising claim that a suitably-defined
``short-time diffusivity'' can actually \emph{increase} with increasing
disorder strength (see Fig. 3 of Ref.\ \cite{Sahimi}). 
Exact bounds, renormalization group
calculations, and previous
numerical simulations are inconsistent with
these results in three dimensions.

At low concentrations of mobile ions, the
 Green function for a diffusing ion 
should obey the diffusion equation:
\begin{equation}
\frac{\partial c_v({\bf r},t)}{\partial t} = 
D_0 \nabla^2 c_v + \beta D_0 \nabla \cdot
[c_v \nabla v({\bf r})] \ .
\label{1}
\end{equation}
Here $c_v$ is the Green function of a single ion in a given realization of
the quenched random potential $v$, $D_0$ is the ``bare,'' short-time 
diffusivity, and $\beta$ is the inverse temperature.
The mean square displacement is given by
$R_v^2(t) = \int d^d {\bf r} \vert {\bf r} \vert^2 c_v({\bf r},t)$.
The observable mean square displacement is given by an average over all
realizations of the disorder:
$R^2(t) = \langle R_v^2(t) \rangle $.
The effective diffusion diffusion  coefficient is defined  in $d$ 
dimensions by
$D = \lim_{t \to \infty} R^2(t)/ (2~ d~ t)$.

Mehrabi and Sahimi model the disorder by 
a quenched Gaussian random potential field.  The statistics of this potential
field are chosen so that they obey bulk charge neutrality:
$\hat \chi_{vv} ({\bf k}) = 
\gamma/[ k^2 (k^2 + \kappa^2)]$.
Here the potential-potential correlation function is 
$\chi_{vv}({\bf r}) = \langle v({\bf 0}) v({\bf r}) \rangle$, $\kappa$ is
an inverse correlation length, and $\gamma$ is a
measure of the density of defects.
The Fourier transform in
$d$ dimensions is given by $\hat f({\bf k}) = \int d^d {\bf r}
f({\bf r}) \exp(i {\bf k} \cdot {\bf r})$.

The single-ion, random diffusion
 model is a well-studied one in statistical physics, and a variety of
exact results are known.  First, there is an exact bound
for the diffusivity in this system in any dimension \cite{deMasi}:
\begin{equation}
\frac{D}{D_0} \ge \exp[- \beta^2 \chi_{vv}(0)] \ .
\label{3}
\end{equation}
Calculating this bound in three dimensions, one finds
$D/D_0 \ge \exp[-\beta^2 \gamma /(4 \pi \kappa)]$.
This result implies that the motion is diffusive in
three dimensions, \emph{i.e.}\ 
$D > 0$. 
The motion is also diffusive at finite ion concentrations, 
since the dynamical exponent is 2 \cite{Deem2}.
 Therefore, the motion should be asymptotically
diffusive in three dimensions.
  Indeed, previous careful simulations by
Dean, Drummond, and Horgan  on related models
have confirmed the bound
 \cite{Dean}.
Moreover, these simulations have shown that Deem and Chandler's
single-ion prediction  \cite{Deem}
\begin{equation}
\frac{D}{D_0} = \exp[- \beta^2 \chi_{vv}(0)/d]
\label{5}
\end{equation}
is accurate to at least moderate disorder strengths.  In fact,
this equation is
correct to second order in $\beta^2 \chi_{vv}(0)$ in all dimensions and is
exact in one dimension.  Note that, as expected
physically,
the diffusion constant
\emph{decreases} with increasing disorder strength.

The situation is more interesting in two dimensions, where
anomalous diffusion can occur
[the bound in Eq.\ (\ref{3}) vanishes]. 
 Indeed, field-theoretic treatments have shown that the
exponent in Eq.\ (\ref{3a}) is continuously variable and
is given exactly by 
$\delta = 1/[ 1+ 8\pi \kappa^2/(\beta^{2} \gamma )]$
\cite{Bouchaud3}.
This scaling has been confirmed by numerical simulations \cite{Victor}.
At finite ion concentrations,
the anomalous diffusion persists at high temperature
 \cite{Deem2},
although the mobile ions may partially screen the disorder.
 A Kosterlitz-Thouless transition
can occur at low temperature \cite{Deem2}.
%In one dimension, the scaling is also exactly known,  and it is given by
%$R^2(t) \sim (\mathrm{const})
% (\ln t)^4$ (see Table 4.2 of Ref.\ \cite{Bouchaud3}).

\end{document}